\documentclass[10pt,emulateapj,apj]{emulateapj}
\shorttitle{Logarithmic halo mass density}
\shortauthors{Jee, et al.}
\begin{document}
\title{A Second-order bias model for the Logarithmic Halo Mass Density}
\author{Inh Jee$^1$, Changbom Park$^2$, Juhan Kim$^3$, Yun-Young Choi$^4$, \& Sungsoo S. Kim$^5$}
\affil{$^1$Department of Astronomy, The University of Texas, 1 University Station, C1400, Austin, Texas 78712-0259\\
$^2$School of Physics, Korea Institute for Advanced Study, Heogiro 85, Seoul 130-722, Korea\\
$^3$Center for Advanced Computation, Korea Institute for Advanced Study, Heogiro 85, Seoul 130-722, Korea\\
$^4$Dept. of Astronomy \& Space Science, Kyung Hee University,
Gyeonggi 446-701, Korea; yy.choi@khu.ac.kr\\
$^5$School of Space Research, Kyung Hee University,
Gyeonggi 446-701, Korea}

\begin{abstract}
We present an analytic model for the local bias of dark matter halos
in a $\Lambda$CDM universe.
The model uses the halo mass density instead of the halo number density
and is searched for various halo mass cuts, smoothing lengths,
and redshift epoches. 
We find that, when the logarithmic density is used,
the second-order polynomial can fit the numerical relation between 
the halo mass distribution and the underlying matter distribution 
extremely well.
In this model the logarithm of the dark matter density is expanded in terms of
log halo mass density to the second order.
The model remains excellent for all
halo mass cuts (from $M_{\rm cut}=3\times10^{11}$ to
$3\times10^{12}h^{-1}M_{\odot}$), 
smoothing scales (from $R=5h^{-1}$Mpc to $50h^{-1}$Mpc),
and redshift ranges (from $z=0$ to 1.0)
considered in this study.
The stochastic term in the relation is found not entirely random,
but a part of the term can be determined by the magnitude of the shear tensor.
\end{abstract}


\section{Introduction}

Understanding the formation and distribution of galaxies
requires knowledge on initial conditions, effects of gravitational evolution,
dark matter halo formation, and galaxy formation in dark halos.
The standard paradigm is to adopt
Gaussian initial density fluctuations that grow through gravitational
instability to form dark matter halos.  Galaxies are thought to form and
evolve in dark halos, which in turn undergo a series of mergers and accretion
(White \& Rees 1978).
It is therefore expected that the observed galaxy distribution is somewhat
different from the halo distribution and also from the underlying
matter distribution even though it is popularly assumed to trace
the dark matter with a constant bias on very large scales.
In many future surveys of galaxy redshifts
it is hoped to measure cosmological parameters with a high precision from
the galaxy distribution, and
it is required to know the relations among matter density field,
distribution of dark matter halos and galaxy distribution very accurately.

Cosmological N-body simulation is a useful tool that can be used to find
the relation between matter and halo distributions,
which we will focus on in this paper.
Manera \& Gaztanaga (2011) recently studied a local halo bias model using
an N-body simulation.  They adopted a non-linear, local, deterministic bias
model of Fry \& Gaztanaga (1993) and Taylor-expanded the halo number
density contrast $n_h$ as a second-order polynomial of the matter fluctuation
$\delta_m$. The coefficients of the linear and quadratic terms are measured
as functions of cubical pixel size and halo mass cut. In this paper we extend
their work to obtain a halo bias model that describes the local halo-matter
density relation in a more universal way. There are many other studies which 
adopted this model ; Guo \& Jing (2009) calculated galaxy bias up to the second 
order from bispectrum based on this model. Roth \& Porciani (2011) applied the 
Eulerian local bias(ELB) model and compared the results between 
Standard Perturbation Theory (SPT) and simulations.  It has proved that SPT 
is a good theoretical model in predicting the bias at fairly large smoothing scale. 
However, since the growth of the structure is not deterministic, it cannot provide 
a full picture in estimating bias. Pollack et al. (2011) used four different 
probes to estimate bias: smoothed density field, power spectra, bispectra and 
reduced bispectra. They claimed that the Fourier-space analysis is more reliable 
since it is free from the mode-mixing problem of the real-space analysis. I
n this study we applied smoothing scale dependent bias to get rid of this problem.

We first adopt to use the halo mass density instead of the halo number density
in the halo bias model. This is because the halo mass density has a much
tighter and simpler relation with the underlying matter density
than the halo number density  (see Fig. 1 in section 3.1 below).
Seljak et al.(2009) reported that weighting central halos by their mass can
significantly reduce stochasticity relative to the dark matter below the
Poisson expectation. Park et al. (2010) showed that the gravitational
shear calculated from the halo mass density (calculated
from the halo distribution by weighting halos with their mass)
has a much tighter relation with the true shear field of dark matter
than that from the halo number density (halos are uniformly weighted
above a certain mass cut).
Observational application of the halo mass-weighting
to local density estimation has been already made by Park et al. (2008),
with an assumption that luminous galaxy mass is proportional to halo mass
(see also Park \& Choi 2009, Park \& Hwang 2009, Hwang \& Park 2009 for
more applications).

Second we use the logarithmic density ln(1+$\delta$) instead of the density
contrast $\delta$ in our halo bias model. It turns out that using the
logarithmic density allows us to find a  very simple analytic bias model
that are good for all smoothing scales, halo mass cuts, and redshifts
considered.
There have been many suggestions for using the logarithmic density
as a model for weakly non-linear density field in cosmology (Cole \& Jones
1991; Colombi 1994; Kayo et al. 2001; Neyrinck et al. 2009).
Cole \& Jones (1991) has proposed to use a lognormal model for
the matter distribution evolved from Gaussian initial conditions.
They provided astrophysical motivations for considering the lognormal model.
In particular, they showed that the mass flow governed by the continuity equation
leads the density field to a lognormal distribution in the non-linear regime
if the velocity fluctuation is assumed linear.
Kayo et al. (2001) showed through a comparison with N-body simulations
that the lognormal probability distribution is a useful empirical model for
the cosmological density fluctuations, which is insensitive to the shape of
the density power spectrum.
Neyrinck et al. (2009) found that nonlinearities in the dark
matter power spectrum are almost absent when the density is transformed to
the log density.
They also provided several reasons to use the logarithmic density mapping.

Third, we expand the matter density in terms of the halo density, which is
opposite to all previous studies. This is required by the simulation data.
The log matter density can be fit excellently by a second-order polynomial
of the log halo mass density, not vice versa. A model of the form
$\delta_m = f(\delta_h)$ is in practice useful when $\delta_h$ is an observed
quantity and $\delta_m$ is the one to be estimated.

As in Manera \& Gaztanaga (2011) we restrict our study of the halo biasing
in the real configuration space. The effects of the redshift-space distortion
will be of practical interest, but an observed distribution of galaxies
in redshift space can be first corrected for the fingers-of-god by shrinking
massive clusters and groups of galaxies and for the large-scale peculiar velocity by
using the second-order perturbation theory (Gramann et al. 1994)
before the halo bias model in real space is applied.
We will briefly discuss the halo bias in Fourier space in section 5.

\section{N-body simulation}
	\subsection{Simulation}
The simulation used in this paper is based on the Wilkinson Microwave
Anisotropy Probe (WMAP) three-year parameters (Spergel et al. 2007),
$\Omega_{\Lambda}=0.762$, $\Omega_{m}=0.238$, $\Omega_{b}=0.042$,
$n_s=0.958$, $h=0.732,$ and $\sigma_8=0.761$, where $\Omega_{\Lambda}$,
$\Omega_{m},$ and $\Omega_{b}$ are the density parameter  associated
with the cosmological constant, matter, and baryon, respectively,
$n_s$ is the slope of the primodial power spectrum, and  $\sigma_{8}$
is the rms fluctuation of the matter density field smoothed with a 8
$h^{-1}$Mpc radius top-hat sphere.
The simulation evolved $2048^3$ cold dark matter particles
with mass $9.6 \times 10^9 h^{-1}M_{\odot}$, and the minimum halo
mass was $2.9 \times 10^{11} h^{-1}M_{\odot}$ (30 particles).
Initial conditions were generated on a $2048^3$ mesh, in accordance with
the $\Lambda$CDM power spectrum (Eisenstein \& Hu 1998).
The physical size of the simulation cube is $1024 h^{-1}$Mpc.
The initial epoch of the simulation
was $z = 47$, and 1880 global time steps were taken until the simulation reaches
the present epoch.
To increase the spatial dynamic range we use a parallel
N-body code made by Dubinski et al. (2003, 2004) which is a merger between
a PM code (Park 1990, 1997) and a tree-code (Barnes \& Hut 1986).

	\subsection{Dark Halos}
Since our study compares the halo distribution with the distribution
of the underlying dark matter, the way halos are defined
can be a crucial problem.
The most widely used method, the friends-of-friend (FoF) algorithm, provides
an easy way to identify virialized halo regions using a particle
linking length parameter (Audit et al. 1998; Davis et al. 1985).
Matching the FoF halos with the observed galaxies is unrealistic because
each FoF halo can contain many galaxies.
A better model is to use subhalos each of which is assumed to contain one
galaxy. In this model galaxy mass or luminosity is assigned to halos with
the constraint that the halos with mass above a certain limit
have the number density equal to that of the galaxies with stellar mass or
luminosity above a certain limit. This one-to-one correspondence model or
abundance matching method of galaxy assignment scheme has been shown
to be quite successful in describing the observed galaxy distribution
(Marinoni \& Hudson 2002; Vale \& Ostriker 2004, 2006; Shankar et al. 2006;
Kim et al. 2008; Gott et al. 2009; Choi et al. 2010).

Below we will show that the matter density field has a much higher correlation
with the halo mass density field than the halo number density field.
If the subhalo mass is weighted to each halo, the mass density
fields of the FoF halos and of the subhalos will be the same and it is not
important which halo identification scheme is used.

To identify subhalos, the physically self-bound (PSB) group-finding algorithm
developed by Kim \& Park (2006) and Kim et al. (2008) is used.
The PSB method first finds local particle groups using the FoF algorithm and convert
the particle distribution in each group into a density field measured at
particle positions using a variable-size Spline kernel.
The particles in each group are assigned into subgroups near the
density peaks. The subgroup particles are used to redefine the subgroup member
particles to identify the tidally stable and gravitationally self-bound subhalos
through iteration.
The PSB group-finding method has its advantage in resolving halos even in dense environment, and
can resolve subhalos down to the gravitational force resolution.
It should be noted that subhalos are not substructures of the host halo,
but immigrated objects that are not yet disrupted within the host and
thus should be resolved and treated separately when individual self-gravitating
non-linear objects are to be identified. After finding the FoF halos using a standard 
linking length, the PSB algorithm is applied to identify subhalos. In each FoF halo 
the most massive subhalo is named the central or main subhalo, and the rest are 
called subhalos. In this study, we hereafter use the term `halo' for the 
both kinds of subhalos.

\section{Matter density-halo mass density relation}
	\subsection{Halo mass density field}
When the spatial clustering of galaxies is studied, traditionally
galaxies are uniformly weighted or weighted only according to
 the selection function. The resulting
number density field is used to calculate clustering amplitude statistics
such as the two-point correlation function and the power spectrum
(Davies \& Peebles 1983; Park et al. 1994).
The uniform weighting is still widely used to study the non-linearity,
scale-dependence, and bias in the distribution of galaxies
(Tegmark et al. 2006).
The main reason for using the uniform weighting is to reduce the shot noise.
If galaxy luminosity is used as weight, for example, the resulting
luminosity density field is dominated by the rare brightest objects.
However, it has been recently shown that the halo mass density field
has a much tighter correlation with the underlying matter field
than the halo number density field (Park et al. 2010), and
the uniform weighting cannot be justified because the bias function of
the galaxy number density field is not only more complicated but also
less correlated with the matter density field compared to the galaxy
mass density field as we show in Figure 1.

In Figure 1 we demonstrate that the halo mass density field has
a tighter relation with the matter density field than the halo number density.
The $y$-axis in the upper panel of Figure 1 is the overdensity in
the halo number density,
and that in the lower panel is the overdensity in the halo mass density.
The density fields are smoothed with a Gaussian filter with radius of
 $R=5 h^{-1}$Mpc.
It can be seen that the halo mass density has a relation
with the underlying matter density not only much tighter but also
much simpler than the halo number density.
The disadvantage of using the number density is even more serious if
the FoF halos are used because the number density of the FoF halos
is relatively low in high density regions.

\begin{figure}
\epsscale{1}
\plotone{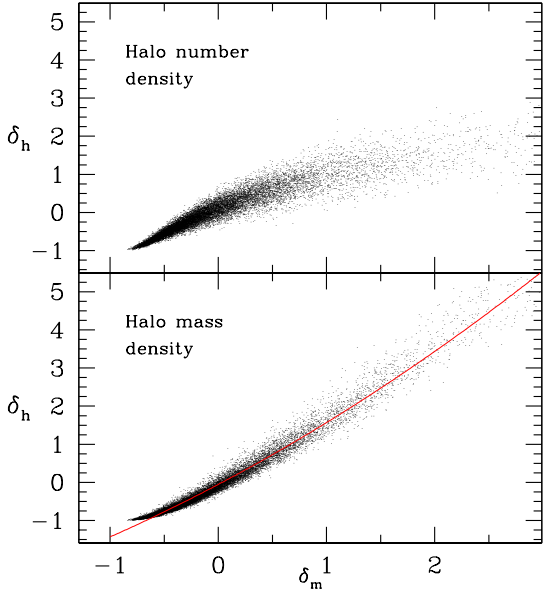}
\caption{The relation of the matter density fluctuation $\delta_m$ with the halo
number density fluctuation (upper panel) and halo mass density fluctuation (lower panel).
The relations are obtained on a Gaussian smoothing scale of $5h^{-1}$Mpc. The solid line
in the lower panel is a second-order polynomial fit to the data.
}
\end{figure}

\begin{figure*}[!!t]
\centering 
\plotone{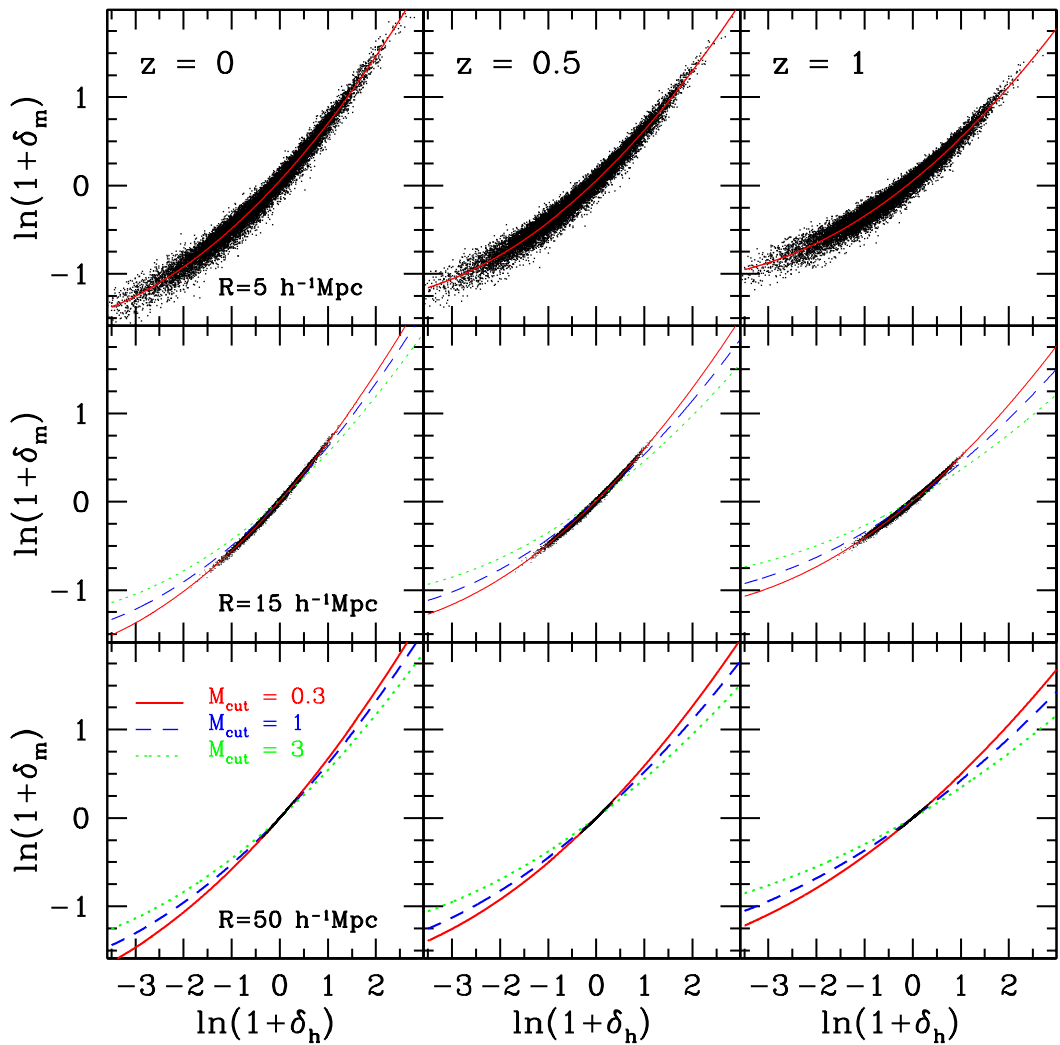}
\caption{The relations between the logarithmic halo mass density and the
underlying logarithmic matter density for
different mass cuts and smoothing scales at different redshifts.
The panels in the first, second and third row are for Gaussian
smoothing length of 5, 15 and 50 $h^{-1}$Mpc, respectively.
Those in the first, second and the third column are the relations
at z=0, 0.5 and 1, respectively. Also shown are the relations for
the cases of three halo mass
cuts of 0.3 (solid line), 1.0 (long-dashed line), and 3.0 (dotted line)
in units of $10^{12}h^{-1}M_{\odot}$ best fit to the data.
}
\end{figure*}

	\subsection{The second-order bias model}

Even though the relation with the matter density becomes much tighter
and simpler when the halo mass density is used, it can not be fit by
a low-order polynomial.
The line in the lower panel of Figure 1 is a second-order polynomial,
$\delta_h =b_0 +b_1 \delta_m +b_2 \delta_m^2$,
best fit to the numerical relation.
It can be seen that the formula can not fit the data at low densities.
We also found that, even if higher-order terms are added,
fitting by a polynomial is not always successful at all redshifts.
This reveals a significant limitation of the bias model of Fry \& Gaztanaga (1993)
working only in a very weak non-linear regime.

We find that, when the logarithmic density is used, a second-order polynomial
relation between the halo mass density and the matter density
stands almost universally. We first smooth the mass-weighted halo distribution
and the matter density field with a Gaussian filter of radius $R$
and take the logarithm of the resulting smooth density fields $D=\ln(1+\delta)$.
The logarithmic transformation can make the result diverge in void regions. 
Since the smoothing length is always chosen to be equal to or greater than 
the mean halo separation, the Gaussian smoothing avoid this situation in practice. 
Note that, when $|\delta| \ll 1$, $D$ approaches $\delta$ and our model reduces
to the conventional linear bias model.
The scatter plots in Figure 2 are the relations between $D_h=\ln(1+\delta_h)$
($x$-axis) and $D_m=\ln(1+\delta_m)$ ($y$-axis) at redshifts $z=0$ (left
column), 0.5 (middle column), and 1.0 (right column). The smoothing scales are
$R=5$ (top row), 15 (middle row), and 50 $h^{-1}$Mpc (bottom row).
Our model for these numerical relations is
\begin{equation}
D_m = \beta_0 +\beta_1 D_h + \beta_2 D_h^2,
\end{equation}
where $\beta_i$'s are functions of the smoothing scale $R$ and
the halo mass cut $M_{\rm{cut}}$. It should be noted that the matter density
is expanded in terms of the halo density. The solid lines in Figure 2 are second-order
polynomials best fit to the scatter plots for the halos with mass more
than $3\times 10^{11} h^{-1}M_{\odot}$. For higher mass cuts of
$M_{\rm{cut}}=1\times 10^{12}$ (long-dashed lines) and
$3\times 10^{12} h^{-1}M_{\odot}$ (dotted lines)
we show only the best fits without scatter plots to avoid confusion.
Since the density fields should be estimated from discrete points 
like halos or galaxies,  the shot noise can be an issue. 
We choose the smoothing length $R$ equal to or greater than the mean 
halo separation to reduce the shot noise. 
As a result for the halo sample with 
$M_{\rm cut}=3\times 10^{11}h^{-1}M_{\odot} ({\bar d}=4.62h^{-1}$Mpc)
we study the halo bias on the scales $R\ge 4.62h^{-1}$Mpc. 
Likewise for a sample with $M_{\rm cut}=1$ or 
$3\times 10^{12}h^{-1}M_{\odot} ({\bar d}=6.56$ or $9.56h^{-1}$Mpc)
the halo bias is studied on scales $R \ge 6.56$ or $9.56h^{-1}$Mpc, 
respectively (see Weinberg et al. 1987 and Park et al. 2005 for choice 
of the smoothing length for the topology study of galaxy distribution).  

\begin{figure*}[!!t]
\centering 
\plotone{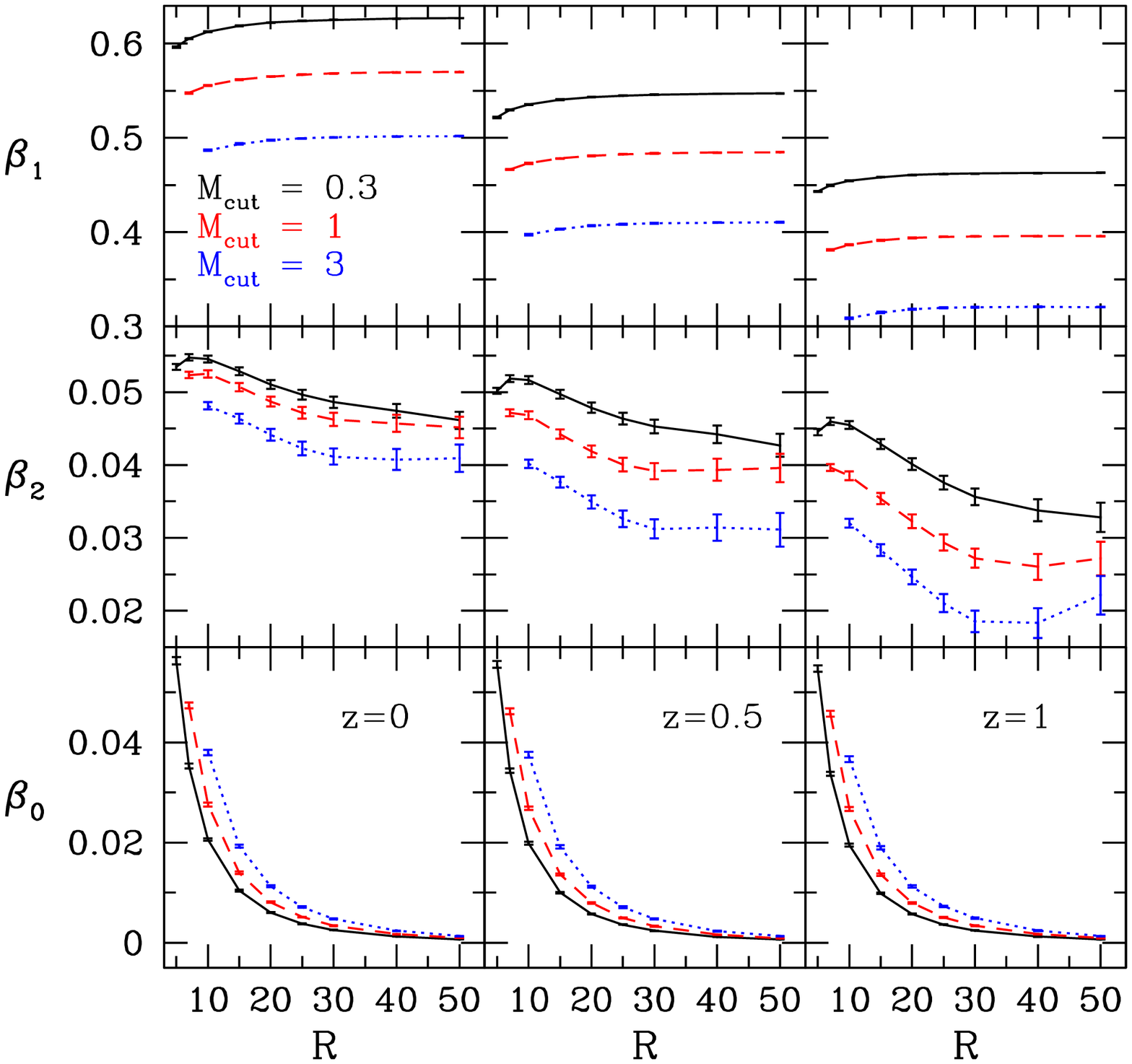}
\caption{
The bias factors $\beta_i$'s in our second-order halo bias model of the
logarithmic density.  For each halo masscut of
$0.3\times10^{12} h^{-1}M_{\odot}$ (solid line), $1\times10^{12} h^{-1}M_{\odot}$
(dashed line) and $3\times10^{12} h^{-1}M_{\odot}$ (dotted line) the bias factors are 
given as a function of the Gaussian smoothing radius.
The bias factors are inspected at three redshift epoches $z=0, 0.5$, and 1.0.
}
\end{figure*}

	\subsection{Dependence on smoothing scale and halo mass cut}

The coefficients $\beta_i$'s in Equation (1) depend on smoothing length
and halo mass cut. Figure 3 shows the dependence of $\beta_i$
on $R$ and $M_{\rm{cut}}$ at three redshifts.
In this paper, all the smoothings are done using the Gaussian filter, 
as stated in section 3.2.
The zero-point offset $\beta_0$ of the halo mass density-matter density
relation is significant on small scales, but rapidly vanishes as $R$ increases.
The zero-point off-set is also seen in the $\delta_m$-$\delta_h$ relation.
On the other hand, the coefficient of the second order term $\beta_2$
does not decrease much as $R$ increases, and Equation (1) remain
quadratic even on very large scales. Equation (1) can be approximated by
a linear bias model $\delta_h = b_1 \delta_m$ on large scales not because
the equation becomes linear but just because the ranges in $D_h$ and $D_m$
become small. This can be seen in Figure 2, where the shape of the fitting
curves hardly changes as $R$ increases.

We fit the values of $\beta_i$'s at $z=0$ over the $R$ ranges from 5 to
$50 h^{-1}$Mpc and the $M_{\rm{cut}}$ range from $0.3\times 10^{12}$ to
$3.0\times 10^{12} h^{-1}M_{\odot}$ using a routine obtained from
http://www.zunzun.com. The functional forms of the coefficients we use are

\begin{eqnarray}
\beta_1  &=&  aM_{\rm cut}^{c} b^{\frac{1}{R}}+d,\\
\beta_2  &=&  \frac{a+b M_{\rm cut}+c{R}+d M_{\rm cut} R}{1+f M_{\rm cut}+
g R+h M_{\rm cut} R},\\
\beta_0 &=&  \frac{a+b M_{\rm cut}+c R+d M_{\rm cut} R}{1+f \ln(M_{\rm cut})+
g \ln(R)+h \ln(M_{\rm cut}) \ln(R)}+i
\end{eqnarray}

Table 1 lists the best-fit values of the constants in these functions
when the halo mass is in units of $10^{12}h^{-1}M_{\odot}$.
The parameter values in this table are useful only for the cosmological 
model we adopted. A simulation using slightly different cosmological 
parameters would give slightly different $\beta$'s. The aim of the 
current study is to show that the mapping from halo mass density to 
underlying matter density can be well-modeled by a second-order 
polynomial when logarithmic densities are used.

\begin{table}
\caption{Coefficients of the second-order bias model at redshift $z=0$.}
\centering
\begin{tabular}{ c|c | c | c | c | c | c | c | c }
\hline
 & $a$ & $b$ & $c$ & $d$ & $f$ & $g$ & $h$ & $i$\\
\hline
\hline
$\beta_1$ & -0.297 & 1.85    & 1.77     & 0.87     &       &          &          &       \\
$\beta_2$ & 0.056  & 7.15E-3 & -6.66E-4 & -7.32E-5 & 0.175 & -9.39E-3 & 1.08E-4  &       \\
$\beta_0$ & 1220   & 11.5    & 14.5     & -0.315   & -1310 & 4490     & 354      & -0.105\\
\hline
\end{tabular}
\label{table:b3}
\end{table}

	\subsection{Dependence on redshift}

\begin{figure}
\epsscale{1}
\plotone{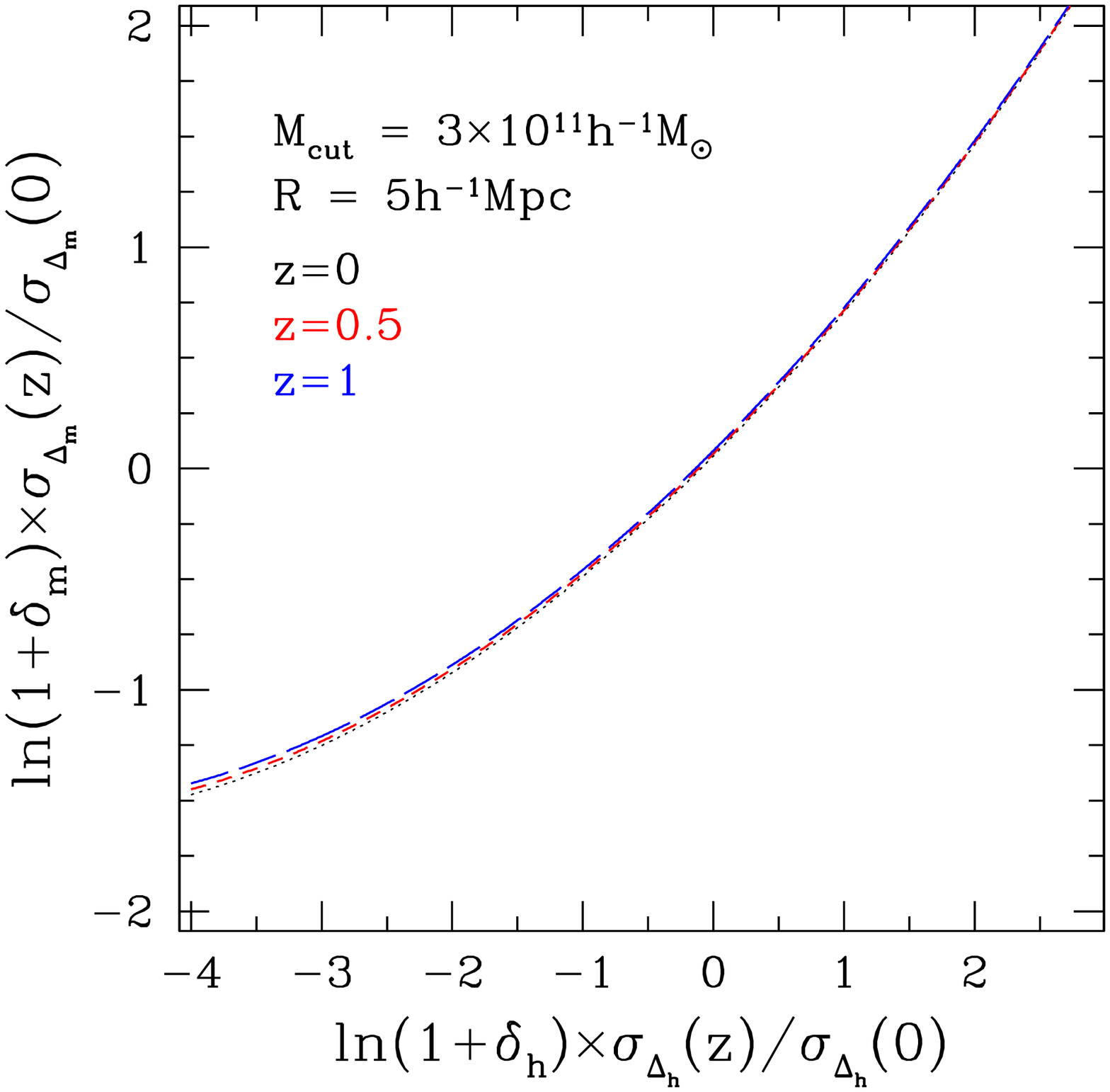}
\caption{
The relation between $D_h=\ln(1+\delta_h)$ and $D_m=\ln(1+\delta_m)$ at redshift
$z=$0 (dotted line), 0.5 (short-dashed) and 1.0 (long-dashed).
The lines for $z=0.5$ and 1 are scaled by $D\sigma(z)/\sigma(0)$.
}
\end{figure}

We find that the redshift dependence of the $D_h$-$D_m$ relation can be accurately
modeled through rms value matching of the $D$ fields across different
redshifts. When a second-order $D_h$-$D_m$ relation is found at one epoch
$z_1$, the corresponding equation at a
different epoch $z_2$ is accurately given if $D$'s are scaled to
$D\sigma_D(z_2)/\sigma_D(z_1)$. This is demonstrated in Figure 4. It shows
the $D_h$-$D_m$ relations for $M_{\rm{cut}}=3\times 10^{11}h^{-1}M_{\odot}$
and $R=5h^{-1}$Mpc at redshifts $z=0$ (dotted line), 0.5 (short-dashed
line), and 1.0 (long-dashed line).
The lines for the $z=0.5$ and 1 cases are drawn by scaling the $x$ and
 $y$-axes by $\sigma_D(z)/\sigma_D(0)$.
It can be seen that the lines are very close to one another once
the scaling is made. To make Equation (1) applicable at different redshifts
we introduce a scaled variable $\Delta =D\sigma_D(z)/\sigma_D(0)$ and
arrive at a modified halo bias model
\begin{equation}
\Delta_m = \beta_0 +\beta_1 \Delta_h + \beta_2 \Delta_h^2,
\end{equation}
where $\Delta_m =D_m\sigma_{D_m}(z)/\sigma_{D_m}(0)$ and
$\Delta_h=D_h\sigma_{D_h}(z)/\sigma_{D_h}(0)$. 
We emphasize again that the matter density is expanded in terms of the
halo mass density in our model.
$\Delta$'s are equal to
$D$'s at $z=0$. It is reassuring that the redshift dependence of $D_h$ versus $D_m$ relation satisfies this scaling relation down to the smoothing scales as small as 5$h^{-1} Mpc$. This scaling property of our second-order polynomial
bias model also makes the model applicable to cases with a different degree
of biasing.

\section{Stochasticity}
At a particular point in space the halo density is not completely fixed
for a given matter density, but can have a range of values as can be seen
in Figure 2. This dispersion in the relation is called stochasticity
(Seljak et al. 2009).
Ue-Li Pen(1998) claimed that on large scales galaxy density variance, 
galaxy-dark matter density bias, and their cross-correlation coefficient 
are the only coefficients that are required in determining the dark matter 
power spectrum, and in mildly nonlinear regime skewness and non-linear 
bias should come into consideration.
Dekel \& Lahav(1999) proposed conditional distribution in estimating random 
fluctuations of galaxy and mass density. They claimed that the scatter 
in this relation arises from the hidden factors related to shot noise, 
galaxy formation etc. 
Somerville et al. (2001) modeled the non-linear stochastic bias as a combination 
of mean biasing function and scatter about it. It was claimed that it depends on 
halo mass, luminosity and scale. They investigated the time dependence of bias 
in detail, and concluded that the evolution of biasing depends largely on the 
cosmological model due to the growth rate.
Seljak \& Warren (2004) estimated the halo bias as approximately constant for halo 
masses one-tenth below the non-linear mass. They also explored the growth of 
individual Fourier modes of dark matter density field and found large fluctuations 
due to non-linearity. Similar large fluctuations were found between modes of 
halos and dark matter densities, and between modes of halo density fields 
with different halo mass. In a succeeding study, Bonoli \& Pen (2009) tested 
the validity of basic assumption of halo bias studies that presumes no 
intrinsic stochasticity between halo and dark matter. From the behavior of 
stochasticity in the correlation between halos with different mass and 
between halos and dark matter as a function of mass, they claimed that even 
on large scales, stochasticity cannot be neglected. 
Roth \& Porciani (2011) tested the third-order standard perturbation theory (SPT) 
by comparing the dark matter field produced by simulation with the 
numerically-calculated dark matter field using the SPT from initial conditions, 
both point-by-point and statistically. They claimed that on large scales above 
8$h^{-1}$Mpc, they agreed well up to redshift 0. To relate the matter field 
to halo density fields identified from the simulation, they applied Eulerian 
bias (ELB) model up to the third order by fitting scatter plots. 
Using this bias they reconstructed the halo field to show that it cannot 
reproduce all the detailed properties of halo distribution. 
On large scales they showed that bias can be approximated as a constant and 
it agrees with the parameter $b_1$ from the local bias model, in those regions 
with $\delta<<1$. As a consequence they concluded that the SPT is a good method 
to study the matter field on fairly large smoothing scales, but the ELB model 
cannot fully reconstruct the halo density field.

Since all the matter in $\Lambda$CDM universes is contained in halos, 
$\Delta_h$ is identical to $\Delta_m$ and the $\Delta_h$-$\Delta_m$
relation will be exactly linear
when the halo mass cut is zero. Since all the matter in the LCDM universes 
is contained in halos in the case of a pure N-body simulation,then the stochastic
term is thought to be caused by the background uncounted halos with mass below the
mass cut. (In practice, the unresolved halos below the mass cut is responsible only
for a part of stochasticity since there can be other forms of mass components whose 
distribution can be governed by various environmental factors.) To reduce the 
stochasticity Hamaus et al. (2010) suggested a simple halo
weighting scheme where the weight to a halo with mass $M_h$ is not $M_h$
but $M_h+M_{\rm{cut}}$.
When we implement this weighting scheme into the density calculation,
we find the $\delta_h$-$\delta_m$ relation slightly changes
and its dispersion somewhat decreases.
We find the reduction of the scatter is about 10\% in the case of
$M_{\rm{cut}}=3\times 10^{11}M_{\odot}$ and $R=5h^{-1}$Mpc. Since
the reduction of stochasticity is not significant, we will not adopt
this weighting.

We look for a possibility that some local parameters other than the
local density have information on the underlying matter density, and
a part of the stochasticity in the $\Delta_h$-$\Delta_m$ relation is 
actually not random but can be determined by such environmental parameters.
We investigate the dependence of the scatter on a few local and
nonlocal physical parameters. They include the gravitational shear
tensor $\partial_i\partial_j \Phi$, Laplacian of the logarithmic halo
density field  $\nabla^2\Delta$, and the difference between the 
densities on two smoothing scales $\Delta(R_1)-\Delta(R_2)$. 
The mass cut is fixed to $M_{\rm{cut}}=3\times 10^{11}h^{-1}M_{\odot}$ 
in this section.

The halo collapse time and the mass accreted till an epoch can depend 
on various environmental parameters, and the halos with mass above
$M_{\rm{cut}}$ located in the same local density regions can have 
different physical parameters and the shape of the mass function can
depend on environment.
For example, the gravitational shear force can influence the halo formation
under the same density environment (McDonald \& Roy 2009);
see also Chan et al. (2012) and Tobias et al. (2012).
We calculate the traceless Hessian matrix of the gravitational potential
\begin{equation}
H_{ij} = \frac{\partial^2\Phi}{\partial x_i \partial x_j}-{1\over3}
\delta_{ij}\nabla^2\Phi
\end{equation}
from the smooth halo mass density and matter density fields. 
The traceless Hessian matrix is used
 to retain only the anisotropic component of the gravitational potential
 field. Its eigenvalues $\lambda_i$ are used to
define the ellipticity and prolateness parameters
\begin{eqnarray}
e &=& (\lambda_1-\lambda_3)/\lambda,\\
p &=& (\lambda_1-2\lambda_2+\lambda_3)/\lambda,
\end{eqnarray}
where $\lambda_1\ge\lambda_2\ge\lambda_3$ is assumed and the magnitude
of the shear tensor $\lambda = \sqrt{\lambda_1^2+\lambda_2^2+\lambda_3^2}$.
The curvature of the gravitational potential field determined the shear tensor, which produces the tidal torque on density fluctuations.
The ellipticity and prolateness quantifies the shape of the potential by comparing
the size of the curvature in directions of the 
principal axes, while the shear magnitude gives the size of
the curvature itself. The ellipticity and prolateness parameters are
divided by the shear magnitude so that only the information
on the potential shape can be obtained.

\begin{figure}
\epsscale{1}
\plotone{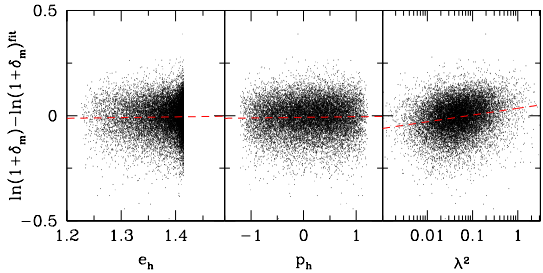}
\caption{
The scatter term of $\ln(1+\delta_m)$ from the second-order halo bias model
as a function of ellipticity, prolateness and shear magnitude
calculated from the halo distribution.
}
\end{figure}

Figure 5 is the scatter term in $\ln(1+\delta_m)$ from the fitting formula
 shown in Figure 2 (the panel at the upper left corner). 
 It can be seen that the scatter term does not depend on the normalized
 ellipticity or prolateness parameters, but weakly depends on the shear
 magnitude. A least-square fit of a line results in
\begin{equation}
\ln(1+\delta_m)-\ln(1+\delta_m)_{\rm fit} = 0.035 + 0.032 \log_{10}(\lambda^2),
\end{equation}
where the error in the slope is $ 0.01$.

\begin{figure}
\epsscale{1}
\plotone{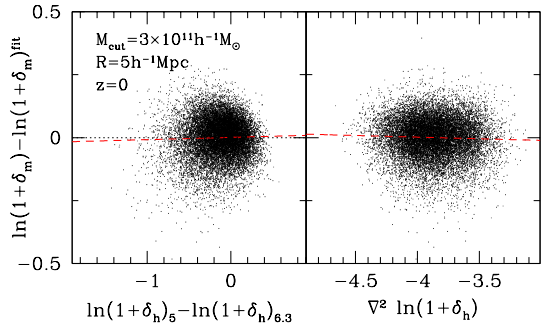}
\caption{
The scatter term from the second-order halo bias model versus
the difference between $\Delta_h$ on two smoothing scales (left panel) 
and the Laplacian of $\Delta_h$ (right panel).
}
\end{figure}

In the left panel of Figure 6 we show the scatter term as a
function of the density difference between $\ln(1+\delta_h)$ with $R=
5 h^{-1}$Mpc and $6.3h^{-1}$Mpc, a measure of non-locality. The smoothing 
volume in the latter case is twice that of the first case. We do not find 
a statistically significant dependence of the scatter on the density 
difference. The right panel shows the scatter term as a function of
the Laplacian of the smooth halo mass density. Again there is no
significant correlation.

Our results can be compared with the theoretical prediction of McDonald
\& Roy (2009) who showed that the galaxy density can be Taylor-expanded
in terms only of $\delta_m$, $\delta_m^2$, and $\lambda_m^2$ up to the 
second order.
To conclude we find that the scatter term in $\ln(1+\delta_m)$ shown in Figure 2
can be partly determined by the magnitude of the gravitational shear
tensor, and the halo bias model can be improved to the following formula
\begin{equation}
\Delta_m = \beta_0 +\beta_1 \Delta_h + \beta_2 \Delta_h^2+d(\lambda),
\end{equation}
where the extra deterministic term $d(\lambda)$ is given by
Equation (9) for the case of $M_{\rm cut}=3\times10^{11}h^{-1}M_{\odot}$
and $R=5h^{-1}$Mpc.
For an observed distribution of dark halos Equation (10) gives the
estimate of the underlying matter density field.

\section{Summary and Discussion}

We present an analytic model for the local bias of dark matter halos
simulated in an N-body simulation of the $\Lambda$CDM universe.
We found that a second-order polynomial model for the relation between
the halo mass distribution and the underlying matter distribution is an
excellent fit to N-body simulation data when the logarithmic density
is used. The model is second-order not in the matter density, but
in the halo mass density.
The model remains excellent for all smoothing scales
(from $R=5h^{-1}$Mpc to $50h^{-1}$Mpc),
halo mass cuts (from $M_{\rm cut}=3\times10^{11}$ to
$3\times10^{12}h^{-1}M_{\odot}$), and redshift ranges (from $z=0$ to 1.0)
considered.
The scatter term in the relation between the halo mass density and
matter density is found not entirely random. We showed that a fraction
of the scatter can be determined by the magnitude of the shear tensor
and the scatter can be reduced.  

Cen \& Ostriker (1993) claimed that the following second-order polynomial 
relation between matter and galaxies
\begin{equation}
\log(\frac{n_g}{\langle n_g\rangle}) = A + B \log(\frac{\rho_{tot}}{\langle\rho_{tot}\rangle})
+ C [\log (\frac{\rho_{tot}}{\langle\rho_{tot}\rangle})]^2
\end{equation}
was an excellent fit to their simulation results of Cen \& Ostriker (1992).
Here $n_g$ is the simulated galaxy number density and $\rho_{tot}$
is the total mass density. It should be noted that Equation (11) is completely
different from our model even though it also adopted a second-order polynomial
of logarithmic density. First, it uses galaxy number density, and it is expected
from Figure 1 that the galaxy number density is not related with the total
mass density through a simple polynomial. Second, Equation (11) is second-order
in log total matter density but our model (Eq. 1) is second-order in the halo mass
density and linear in the matter density. We checked Figure 4 of
Cen \& Ostriker (1992) and found the figure in actual fact supports our model
instead of Equation (11).

Manera \& Gaztanaga (2011) studied a local bias model of the halo distribution
in N-body simulations. However, they also used the halo number density 
instead of the mass density, and adopted a polynomial model where the halo 
overdensity is expanded in terms of the matter overdensity up to the second order.
The mean separation of the FoF halos they used was $11 h^{-1}$Mpc even in the lowest mass halo sample at $z=0$, and their study was reliable only on large smoothing scales of $R\ge 11 h^{-1}$Mpc (or cubical cell size $\ge 28h^{-1}$Mpc). Therefore the data was not quite appropriate for the halo bias study in the non-linear regime. Furthermore, the cubical cell smoothing kernel they used produced a large shot noise in the halo density, which coupled with the true smoothing length dependence of the bias factors.

\begin{figure}
\epsscale{1}
\plotone{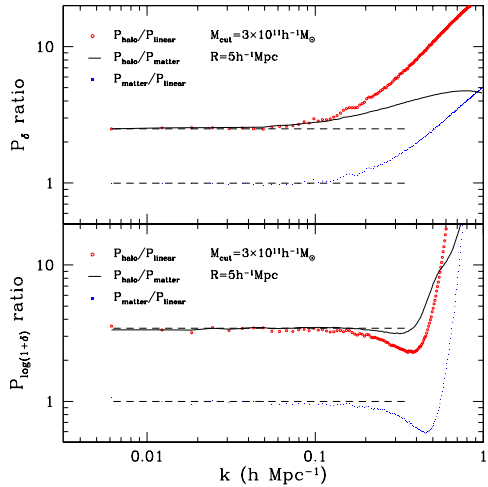}
\caption{
The upper pannel is the ratio of the power spectrum calculated from the 
$\delta$ fields, and the lower pannel is the same plot but
calculated from the ln(1+$\delta$)field, in Fourier space. Red dots indicate 
the ratio between the halo power spectrum and matter power spectrum. 
Blue dots are the ratio between the matter and initial (Gaussian random field) 
power spectrum to show how the evolved matter field is deviated from the
initial power spectrum. 
The black solid lines are the ratios between the halo and matter
power spectra that show the scale-dependent halo biasing in Fourier space.
The horizontal lines are constant bias relations.
}
\end{figure}

Recently, Neyrinck et al. (2009, 2011) found that the linear regime in the 
dark matter power spectrum can be dramatically extended to high $k$'s
if a logarithmic mapping or a Gaussianization for the density field is made.
This is a very useful finding because the size of observational data
used for primordial parameter estimation effectively becomes many orders 
of magnitude larger by simply making such a trivial non-linear transformation.
We do not inspect the relation between the matter and halo densities
in Fourier space in detail in this paper, but just demonstrate that 
the linear regime extends to much smaller scales if the logarithmic
transformation is made. Neyrinck et al. (2009) showed the same for
the dark matter field, but we show that the same is true even if
the halo mass density field is used.

Each N-body simulation gives the dark matter density down to very small scales,
and the log density can be calculated. But in practice, galaxies and dark halos
are much sparser and their spatial distributions should be smoothed 
to give smooth density fields. To obtain a reliable density estimation
we adopt to use a Gaussian smoothing length equal to or greater than the
mean halo separation.
Figure 7 shows the behavior of the halo bias in terms of the power spectrum
of the smoothed halo mass density fields.
The black solid line in the upper panel indicates that the halo mass power spectrum 
($P_{\rm halo} = \langle|\delta_h^2|\rangle$) deviates from the linear
relation with the dark matter power spectrum $P_{\rm matter}$ 
starting from $k\approx 0.05 h {\rm Mpc}^{-1}$. However, this scale becomes
about $k\approx 0.2 h {\rm Mpc}^{-1}$ as shown in the lower panel
when logarithmic values are used.
The scale of the nonlinear halo bias is effectively reduced by a factor of 
about 4. We have calculated this halo bias function using another simulation
with the WMAP 5-year cosmological parameters and different random initial
conditions, and found this conclusion remains valid.
It is expected that the linear-bias regime can be extended to even smaller 
scales if one can apply a smaller smoothing for a denser sample.

Gaussianization has a similar effect on the density field
(Neyrinck et al.  2009, 2011).
It should be noted that Gaussianization has long been used for the study
of the large-scale topology of galaxy distribution in order to examine
the primordial non-Gaussianity (Weinberg et al. 1987; Choi et al. 2010).

\acknowledgments
This work was supported by a grant from the Kyung Hee University in 2011 (KHU-20100179).
We thank Korea Institute for Advanced Study for providing computing resources
(KIAS Center for Advanced Computation Linux Cluster System) for this work.

{}
\end{document}